\pdfoutput=1

\documentclass[11pt,twoside,a4paper,cmspaper,final,collab]{cms-tdr}
\def\svnVersion{218722}\def\cmsCernNo{2013-075}\def\cmsCernDate{\today}\def\cmsMessage{Published in the European Physical Journal C as \href{http://dx.doi.org/10.1140/epjc/s10052-013-2610-8}{\doi{10.1140/epjc/s10052-013-2610-8.}}}

\begin{document}\cmsNoteHeader{SMP-12-005}

\hyphenation{had-ron-i-za-tion}
\hyphenation{cal-or-i-me-ter}
\hyphenation{de-vices}
\RCS$Revision: 206548 $
\RCS$HeadURL: svn+ssh://svn.cern.ch/reps/tdr2/papers/SMP-12-005/trunk/SMP-12-005.tex $
\RCS$Id: SMP-12-005.tex 206548 2013-09-10 15:57:23Z rocio $
\newcommand{\dyll}{\ensuremath{\cPZ/\Pgg^*\to\ell^+\ell^-}\xspace}
\newcommand{\dytt}{\ensuremath{\cPZ/\Pgg^*\to\Pgt^+\Pgt^-}\xspace}
\newcommand{\dymm}{\ensuremath{\cPZ/\Pgg^*\to\Pgmp\Pgmm}\xspace}
\newcommand{\dyee}{\ensuremath{\cPZ/\Pgg^*\to\Pep\Pem}\xspace}
\providecommand{\re}{\ensuremath{\cmsSymbolFace{e}}} 
\ifthenelse{\boolean{cms@external}}{\providecommand{\cmsLeft}{top}}{\providecommand{\cmsLeft}{left}}
\ifthenelse{\boolean{cms@external}}{\providecommand{\cmsRight}{bottom}}{\providecommand{\cmsRight}{right}}
\cmsNoteHeader{SMP-12-005}
\titlerunning{$\PWp\PWm$ cross section in $\Pp\Pp$ at $\sqrt{s}$ = 7\TeV and limits on anomalous couplings}
\title{Measurement of the $\PWp\PWm$ cross section in $\Pp\Pp$ collisions at $\sqrt{s}$ = 7\TeV and limits on anomalous $\PW\PW\Pgg$ and $\PW\PW\cPZ$ couplings}

\date{\today}

\abstract{
A measurement of $\PWp\PWm$ production
in $\Pp\Pp$ collisions at $\sqrt{s} = 7\TeV$ is presented.  The data were collected with
the CMS detector at the LHC, and correspond to an integrated luminosity of $4.92 \pm
0.11$\fbinv. The $\PWp\PWm$ candidates
consist of two oppositely charged leptons, electrons or muons, accompanied by
large missing transverse energy. The $\PWp\PWm$
production cross section is measured to be $52.4 \pm 2.0\stat \pm 4.5\syst \pm 1.2\lum\unit{pb}$.
This measurement is consistent with the standard model prediction of $47.0 \pm 2.0\unit{pb}$
at next-to-leading order. Stringent limits on the $\PW\PW\Pgg$ and
$\PW\PW\cPZ$ anomalous triple gauge-boson couplings are set.}

\hypersetup{%
pdfauthor={CMS Collaboration},%
pdftitle={Measurement of the W+W- cross section in pp collisions at sqrt(s) = 7 TeV and limits on anomalous WW gamma and WWZ couplings},%
pdfsubject={CMS},%
pdfkeywords={CMS, physics, W-pair production, triple-gauge coupling}}

\maketitle

\section{Introduction}
\label{sec:introduction}
\begin{figure}[h!]
\begin{center}
 \includegraphics[width=0.38\textwidth]{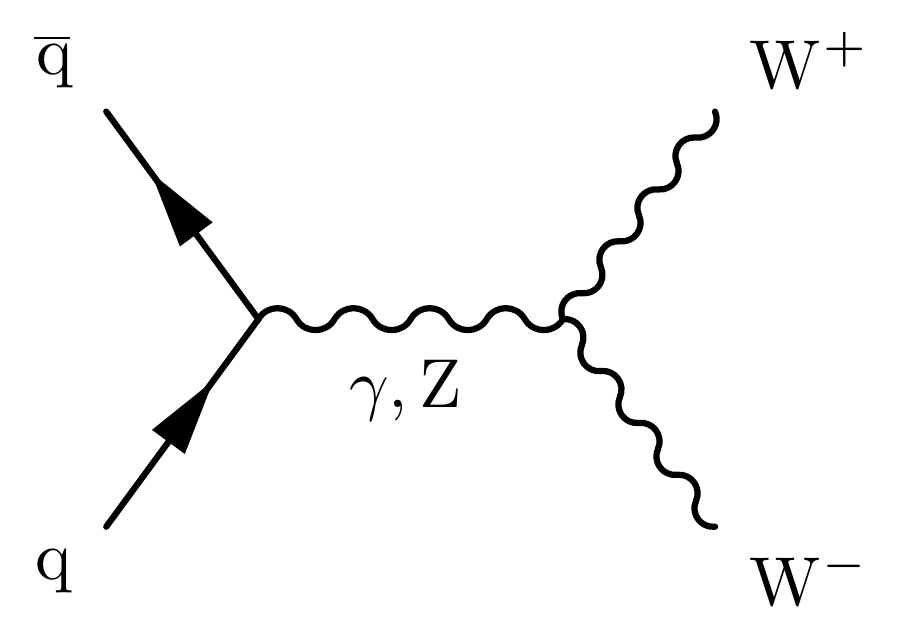}
 \includegraphics[width=0.38\textwidth]{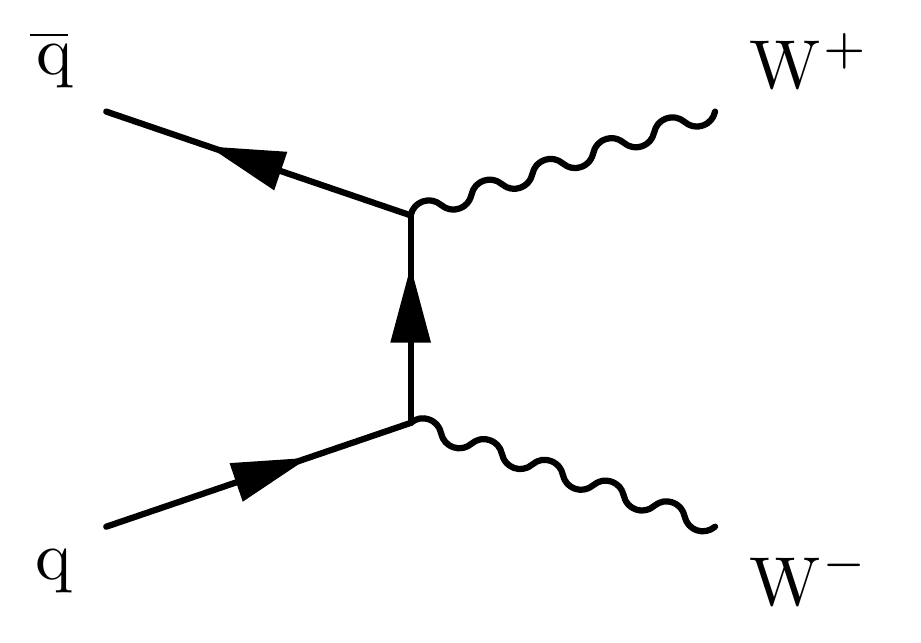}
\caption{Leading order Feynman diagrams for $\cPq\cPaq$ annihilation,
for $s$-channel (\cmsLeft) and $t$-channel (\cmsRight) production of W pairs. The triple
gauge-boson vertex corresponds to the $\PW\PW\Pgg$($\cPZ$) interaction in the
first diagram.}
\label{fig:diagrams}
\end{center}
\end{figure}

The standard model (SM) description of electroweak and strong interactions
can be tested through measurements of the $\PWp\PWm$ production cross section at
a hadron collider. The $s$-channel and $t$-channel $\cPq\cPaq$
annihilation diagrams, shown in Fig.~\ref{fig:diagrams}, correspond to the
dominant process in the SM, at present energies. The gluon-gluon diagrams,
which contain a loop at lowest order, contribute only 3\% of the total cross
section~\cite{MCFM} at $\sqrt{s} = 7\TeV$. $\PW\PW\Pgg$ and $\PW\PW\cPZ$ triple gauge-boson couplings
(TGCs)~\cite{LEPparametrization}, responsible for $s$-channel $\PWp\PWm$ production,
are sensitive to possible new physics processes at a higher mass scale. Anomalous
values of the TGCs would change the $\PWp\PWm$ production rate and potentially certain
kinematic distributions from the SM prediction. Aside from tests of the SM, $\PWp\PWm$
production represents an important background source for new particle searches,
\eg for Higgs boson searches~\cite{HiggsPAS2011,HiggsPaper2012,ATLASHiggsPaper2012}.
Next-to-leading-order (NLO) calculations of $\PWp\PWm$ production in $\Pp\Pp$ collisions at
$\sqrt{s} = 7\TeV$ predict a cross section of $\sigma^\mathrm{NLO} (\Pp\Pp\to\PWp\PWm) = 47.0 \pm 2.0\unit{pb}$~\cite{MCFM}.

This paper reports a measurement of the $\PWp\PWm$ cross section in the
$\PWp\PWm\to\ell^{+}\nu\ell^{-}\cPagn$ final state in $\Pp\Pp$ collisions at
$\sqrt{s} = 7\TeV$ and constraints on anomalous triple gauge-boson couplings.
The measurement is performed with the Compact Muon Solenoid (CMS) detector at
the Large Hadron Collider (LHC) using the full 2011 data sample, corresponding to an integrated
luminosity of $4.92 \pm 0.11$\fbinv, more than two orders of magnitude larger than data
used in the first measurements with the CMS~\cite{Chatrchyan:2011tz} and ATLAS~\cite{atlaswwpaper2010}
experiments at the LHC, and comparable in size to the data sets more recently analysed by ATLAS~\cite{atlaswwpaper2011,atlaswwpaper2012}.

\section{The CMS detector and simulations}
\label{sec:cms}
The CMS detector is described in detail elsewhere~\cite{CMSdetector} so only the
key components for this analysis are summarised here. A superconducting solenoid
occupies the central region of the CMS detector, providing an axial magnetic
field of 3.8\unit{T} parallel to the beam direction. A silicon pixel and strip tracker,
a crystal electromagnetic calorimeter, and a brass/scintillator hadron calorimeter
are located within the solenoid. A quartz-fiber Cherenkov calorimeter extends the
coverage to $\abs{\eta} < 5.0$, where pseudorapidity is defined as $\eta = -\ln[\tan{(\theta/2)}]$,
and $\theta$ is the polar angle of the particle trajectory with respect to the
anticlockwise-beam direction. Muons are measured in gas-ionization detectors embedded
in the steel magnetic-flux-return yoke outside the solenoid. The first level of the
CMS trigger system, composed of custom hardware processors, is designed to select the
most interesting events in less than $3\mus$ using information from the calorimeters
and muon detectors. The high-level trigger processor farm further decreases the rate
of stored events to a few hundred hertz for subsequent analysis.

This measurement exploits $\PWp\PWm$ pairs in which both bosons decay leptonically,
yielding an experimental signature of two isolated, high transverse momentum ($\pt$),
oppositely charged leptons (electrons or muons) and large missing transverse energy
($\MET$) due to the undetected neutrinos. The $\MET$ is defined as the modulus of the
vectorial sum of the transverse momenta of all reconstructed particles, charged and
neutral, in the event. This variable, together with the full event selection, is
explained in detail in Section~\ref{sec:ww_evtsel}.

Several SM processes constitute backgrounds for the $\PWp\PWm$ sample. These include
$\PW+\text{jets}$ and quantum chromodynamics (QCD) multijet events where at least one
of the jets is misidentified as a lepton, top-quark production ($\ttbar$ and $\cPqt\PW$),
Drell--Yan $\dyll$, and diboson production ($\PW\Pgg^{(*)}$, $\PW\cPZ$, and $\cPZ\cPZ$) processes.

A number of Monte Carlo (MC) event generators are used to simulate the signal and backgrounds.
The $\cPq\cPaq\to\PWp\PWm$ signal, $\PW+\text{jets}$, $\PW\cPZ$, and $\PW\gamma^{(*)}$ processes are
generated using the \MADGRAPH 5.1.3~\cite{madgraph5} event generator. The $\cPg\cPg
\to \PWp\PWm$ signal component is simulated using \textsc{gg2ww}~\cite{ggww}. The \POWHEG
2.0 program~\cite{powheg} provides event samples for the Drell--Yan, $\ttbar$, and $\cPqt\PW$ processes.
The remaining background processes are simulated using \PYTHIA 6.424~\cite{pythia}.

The default set of parton distribution functions (PDFs) used to produce the LO MC samples
is CTEQ6L~\cite{cteq66}, while CT10~\cite{ct10} is used for NLO generators.
The NLO calculations are used for background cross sections. For all processes, the detector
response is simulated using a detailed description of the CMS detector, based on the \GEANTfour
package~\cite{Agostinelli:2002hh}.

The simulated samples include the effects of multiple $\Pp\Pp$ interactions in each beam crossing
(pileup), and are reweighted to match the pileup distribution as measured in data.

\section{Event selection}
\label{sec:ww_evtsel}
This measurement considers signal candidates in three final states: $\Pep\Pem$, $\Pgmp\Pgmm$,
and $\Pe^\pm\Pgm^\mp$. The $\PW\to\ell\nu_{\ell}$ ($\ell = \Pe$ or $\mu$) decays are the main signal
components; $\PW\to\tau\nu_{\tau}$ events with leptonic $\tau$ decays are included, although
the analysis is not optimised for this final state. The trigger requires the presence of one
or two high-$\pt$ electrons or muons. For single lepton triggers the $\pt$ threshold for the
selection is 27 (15)\GeV for electrons (muons). For double lepton triggers, the $\pt$ thresholds,
for pairs of leptons of the same flavour, are lowered to 18 and 8\GeV for the first and second
electrons respectively, and to 7\GeV for the each of the two muons. Different flavour lepton
triggers are also used. The overall trigger efficiency for signal events is measured to be
approximately 98\% using data.

Two oppositely charged lepton candidates are required, both with $\pt > 20\GeV$. Electron
candidates are selected using a multivariate approach that exploits correlations between
the selection variables described in Ref.~\cite{egmpas} to improve identification performance,
while muon candidates~\cite{muonpas} are identified using a selection close to that described
in Ref.~\cite{Chatrchyan:2011tz}.  Charged leptons from $\PW$ boson decays are expected to be isolated from any other activity in
the event. The lepton candidates are required to be consistent with
originating at the primary vertex of the event, which is chosen as the vertex with the highest
$\sum \pt^2$ of its associated tracks. This criterion provides the correct assignment for the
primary vertex in more than 99\% of events for the pileup distribution
observed in the data. The efficiency is measured by checking how often a primary vertex with the highest $\sum P_{T}^{2}$ of the constituent tracks is consistent with the vertex formed by the two primary leptons. This is done in MC and checked in data.

The particle-flow (PF) technique~\cite{PFT-09-001} that combines the information from all CMS
subdetectors to reconstruct each individual particle is used to calculate the isolation variable.
For each lepton candidate, a cone around the lepton direction at the event vertex is reconstructed,
defined as $\Delta R = \sqrt{(\Delta\eta)^{2} + (\Delta\phi)^{2}}$, where $\Delta\eta$ and $\Delta\phi$
are the distances from the lepton track in $\eta$ and azimuthal angle, $\phi$ (in radians), respectively;
$\Delta R$ takes a value of 0.4 (0.3) for electrons (muons). The scalar sum of the transverse
momentum is calculated for the particles reconstructed with the PF algorithm that are contained
within the cone, excluding the contribution from the lepton candidate itself. If this sum
exceeds approximately 10\% of the candidate $\pt$, the lepton is rejected; the exact requirement
depends on the lepton flavour and on $\eta$.

Jets are reconstructed from calorimeter and tracker information using the PF technique~\cite{jetpas}.
The anti-\kt clustering algorithm~\cite{antikt} with a distance parameter of 0.5, as
implemented in the \textsc{FastJet} package~\cite{Cacciari:fastjet1,Cacciari:fastjet2}, is used.
To correct for the contribution to the jet energy from pileup, a median energy density $\rho$,
or energy per area of jet, is determined event by event. The pileup contribution to the jet energy is estimated as the
product of $\rho$ and the area of the jet and subsequently subtracted~\cite{Cacciari:subtraction}
from the jet transverse energy $\ET$. Jet energy corrections are also applied as a function of
the jet $\ET$ and $\eta$~\cite{cmsJEC}. To reduce the background from top-quark decays, a jet veto
is applied: events with one or more jets with corrected $\ET > 30\GeV$ and $\abs{\eta} < 5.0$ are rejected.

To further suppress the top-quark background, two top-quark tagging techniques based on soft-muon
and b-jet tagging~\cite{btagpaper,btag2} are applied. The first method vetoes events containing
muons from b-quark decays, which can be either low-$\pt$ muons or nonisolated high-$\pt$ muons.
The second method uses information from tracks with large impact parameter within jets, and
applies a veto on those with the b-jet tagging value above the selected veto threshold. The
combined rejection efficiency for these tagging techniques, in the case of $\ttbar$ events,
is about a factor of two, once the full event selection is applied.

The Drell--Yan background has a production cross section some orders of magnitude larger than the
$\PWp\PWm$ process. To eliminate Drell--Yan events, two different $\MET$ vectors are used~\cite{cmsMET}. The first is
reconstructed using the particle-flow algorithm, while the second uses only the charged-particle
candidates associated with the primary vertex and is therefore less sensitive to pileup. The \textit{projected} $\MET$ is defined as the component of $\MET$ transverse to the direction of the nearest
lepton, if it is closer than $\pi/2$ in azimuthal angle, and the full $\MET$ otherwise. A lower cut
on this observable efficiently rejects $\dytt$ background events, in which the $\MET$ is preferentially
aligned with leptons, as well as $\dyll$ events with mismeasured $\MET$ associated with poorly reconstructed
leptons or jets. The minimum of the projections of the two $\MET$ vectors is used, exploiting the
correlation between them in events with significant genuine $\MET$, as in the signal, and the
lack of correlation otherwise, as in Drell--Yan events.
The requirement for this variable in the $\Pep\Pem$ and $\Pgmp\Pgmm$ final states is \textit{projected} $\MET >
( 37 + N_\text{vtx}/ 2 )\GeV$, which depends on the number of reconstructed primary vertices
($N_\text{vtx}$). In this way the dependence of the Drell--Yan background on pileup is minimised.
For the $\Pe^\pm\Pgm^\mp$  final state, which has smaller contamination from $\dyll$ decays, the
threshold is lowered to $20\GeV$. These requirements remove more than 99\% of the Drell--Yan
background, the actual number of accepted background events is obtained from the data, as explained below.

Remaining  $\dyll$ events in which the Z boson recoils against a jet are reduced by requiring the angle
in the transverse plane between the dilepton system and the most energetic jet to be smaller than 165
degrees. This selection is applied only in the $\Pep\Pem$ and $\Pgmp\Pgmm$ final states when the leading jet
has $\ET > 15\GeV$.

To further reduce the Drell--Yan background in the $\Pep\Pem$ and $\Pgmp\Pgmm$ final states, events with a
dilepton mass within ${\pm}15\GeV$ of the $\cPZ$ mass are rejected. Events with dilepton masses below
$20\GeV$ are also rejected to suppress contributions from low-mass resonances. The same requirement,
where the threshold is lowered to $12\GeV$, is also applied in the $\Pe^\pm\Pgm^\mp$ final state. Finally,
the transverse momentum of the dilepton system ($\pt^{\ell\ell}$) is required to be above $45\GeV$
to reduce both the Drell--Yan background and the contribution from misidentified leptons.

To reduce the background from other diboson processes, such as $\PW\cPZ$ or $\cPZ\cPZ$ production, any event
that has an additional third lepton with $\pt > 10\GeV$ passing the identification and isolation
requirements is rejected. $\PW\Pgg^{(*)}$ background, in which the photon is misidentified as an electron,
is suppressed by stringent $\Pgg$ conversion rejection requirements~\cite{egmpas}.

\section{Estimation of backgrounds}
\label{sec:backgrounds}
A combination of techniques is used to determine the contributions from backgrounds
that remain after the $\PWp\PWm$ selection. The major contribution at this level comes
from the top-quark processes, followed by the $\PW+\text{jets}$ background.

The normalisation of the top-quark background is estimated from data by counting
top-quark-tagged events, with the requirements explained in Section~\ref{sec:ww_evtsel},
and applying the corresponding tagging efficiency. The top-quark tagging efficiency
($\epsilon_\text{top tagged}$) is measured in a data sample, dominated by $\ttbar$ and
$\cPqt\PW$ events, that is selected from a phase space close to that for $\PWp\PWm$ events, but
instead requiring one jet with $\ET >30\GeV$. The residual number of top-quark events
($N_\text{not tagged}$) in the signal sample is given by
\begin{align*}
N_\text{not tagged} = N_\text{tagged} \times (1-\epsilon_\text{top tagged})/\epsilon_\text{top tagged},
\end{align*}
where $N_\text{tagged}$ is the number of tagged events. The total uncertainty on this
background estimation is about 18\%. The main contribution comes from the statistical
and systematic uncertainties related to the measurement of $\epsilon_\text{top tagged}$.

The $\PW+\text{jets}$ and QCD multijet background with jets misidentified as leptons are
estimated by counting the number of events containing one lepton that satisfies the
nominal selection criteria and another lepton that satisfies relaxed requirements on
impact parameter and isolation but not the nominal criteria. This sample, enriched in
$\PW+\text{jets}$ events, is extrapolated to the signal region using the efficiencies for such
loosely identified leptons to pass the tight selection. These efficiencies are measured
in data using multijet events and are parametrised as functions of the $\pt$ and $\eta$
of the lepton candidate. QCD backgrounds are found to be negligible. The systematic
uncertainties stemming from this efficiency determination dominate the overall uncertainty,
which is estimated to be about 36\%. The main contribution to this uncertainty comes from the
differences in the $\pt$ spectrum of the jets in the measurement data sample, composed
mainly of QCD events, compared to the sample, primarily $\PW+\text{jets}$, from which the extrapolation
is performed.

The residual Drell--Yan contribution to the $\Pep\Pem$ and $\Pgmp\Pgmm$ final states outside
of the Z boson mass window ($N_\text{out}^{\ell\ell,\text{exp}}$) is estimated by normalising
the simulation to the observed number of events inside the Z boson mass window in data
($N_\text{in}^{\ell\ell}$). The contribution in this region from other processes where
the two leptons do not come from a Z boson ($N_\text{in}^{\text{non-\cPZ}}$) is subtracted
before performing the normalisation. This contribution is estimated on the basis of the number
of $\Pe^\pm\Pgm^\mp$ data events within the Z boson mass window. The $\PW\cPZ$ and $\cPZ\cPZ$ contributions
in the $\cPZ$ mass window ($N_\text{in}^{\cPZ\mathrm{V}}$) are also subtracted, using simulation, when leptons
come from the same $\cPZ$ boson as in the case of the Drell--Yan production. The residual
background in the $\PWp\PWm$ data outside the Z boson mass window is thus expressed as
\ifthenelse{\boolean{cms@external}}{
\begin{align*}
N_\text{out}^{\ell\ell,\,\text{exp}} &= R^{\ell\ell}_\text{out/in}(N_\text{in}^{\ell\ell} - N_\text{in}^{\text{non-Z}} - N_\text{in}^{ \cPZ\mathrm{V}}),\\
\intertext{ with }R^{\ell\ell}_\text{out/in} &= N_\text{out}^{\ell\ell,\mathrm{MC}}/N_\text{in}^{\ell\ell,\mathrm{MC}}.
\label{eq:dyest}
\end{align*}
}{
\begin{equation*}
N_\text{out}^{\ell\ell,\,\text{exp}} = R^{\ell\ell}_\text{out/in}(N_\text{in}^{\ell\ell} - N_\text{in}^{\text{non-Z}} - N_\text{in}^{ \cPZ\mathrm{V}}),
\text{ with }R^{\ell\ell}_\text{out/in} = N_\text{out}^{\ell\ell,\mathrm{MC}}/N_\text{in}^{\ell\ell,\mathrm{MC}}.
\label{eq:dyest}
\end{equation*}
}
The systematic uncertainty in the final Drell--Yan estimate is derived from the dependence of
$R^{\ell\ell}_\text{out/in}$ on the value of the $\MET$ requirement.

Finally, a control sample with three reconstructed leptons is defined to rescale the
estimate, based on the simulation, of the background $\PW\Pgg^{*}$ contribution
coming from asymmetric $\Pgg^{*}$ decays, where one lepton escapes detection~\cite{wgammastart}.

Other backgrounds are estimated from simulation. The $\PW\Pgg$ background estimate
is cross-checked in data using the events passing all the selection requirements
except that the two leptons must have the same charge; this sample is dominated by
$\PW+\text{jets}$ and $\PW\Pgg$ events. The  $\dytt$ contamination is also cross-checked using
$\dyee$ and $\dymm$ events selected in data, where the leptons are replaced with
simulated $\Pgt$-lepton decays, and the results are consistent with the simulation.
Other minor backgrounds are $\PW\cPZ$ and $\cPZ\cPZ$ diboson production where the two selected
leptons come from different bosons.

\begin{figure*}[htbp]
\begin{center}
 \includegraphics[width=0.48\textwidth]{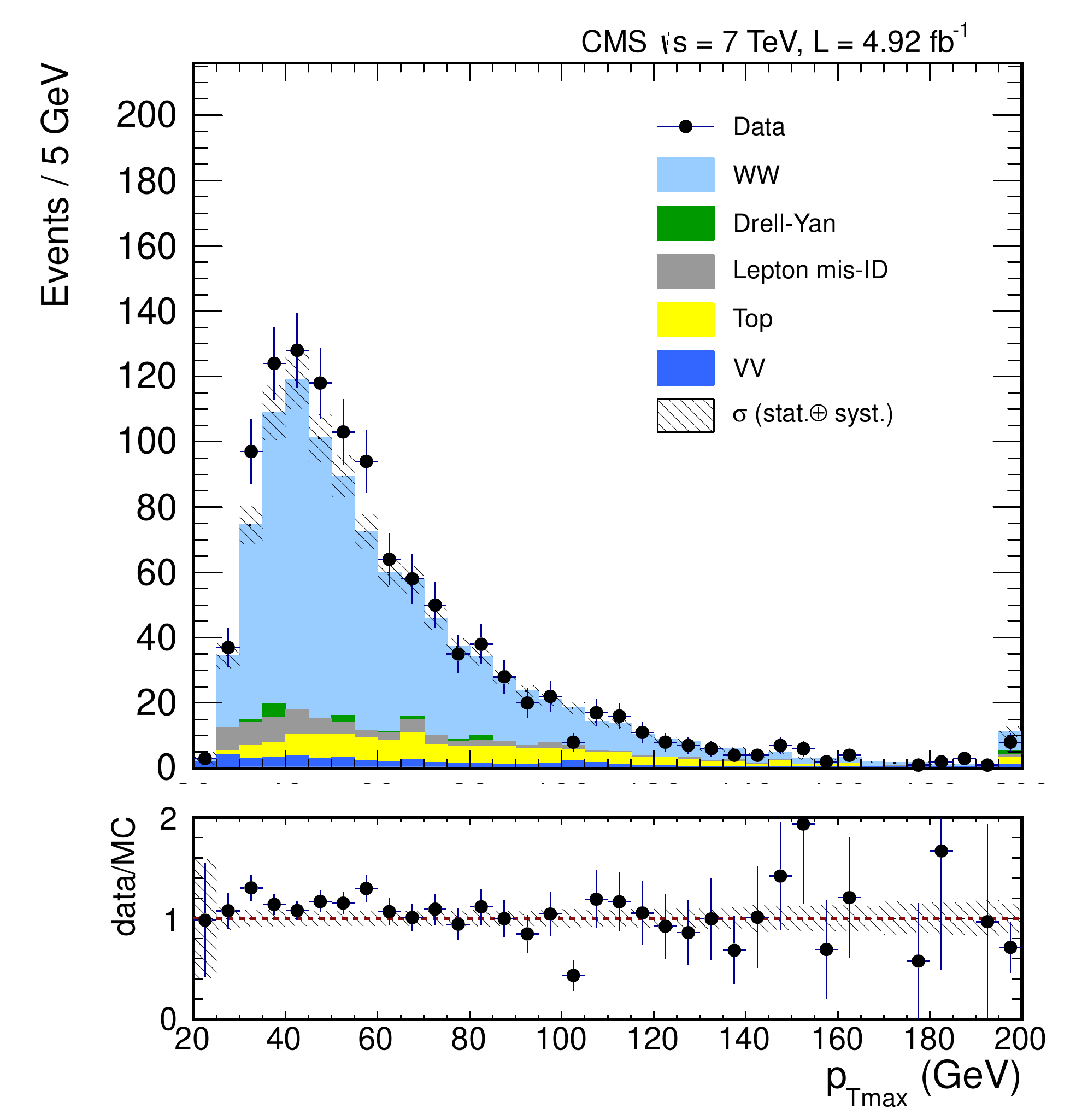}
 \includegraphics[width=0.48\textwidth]{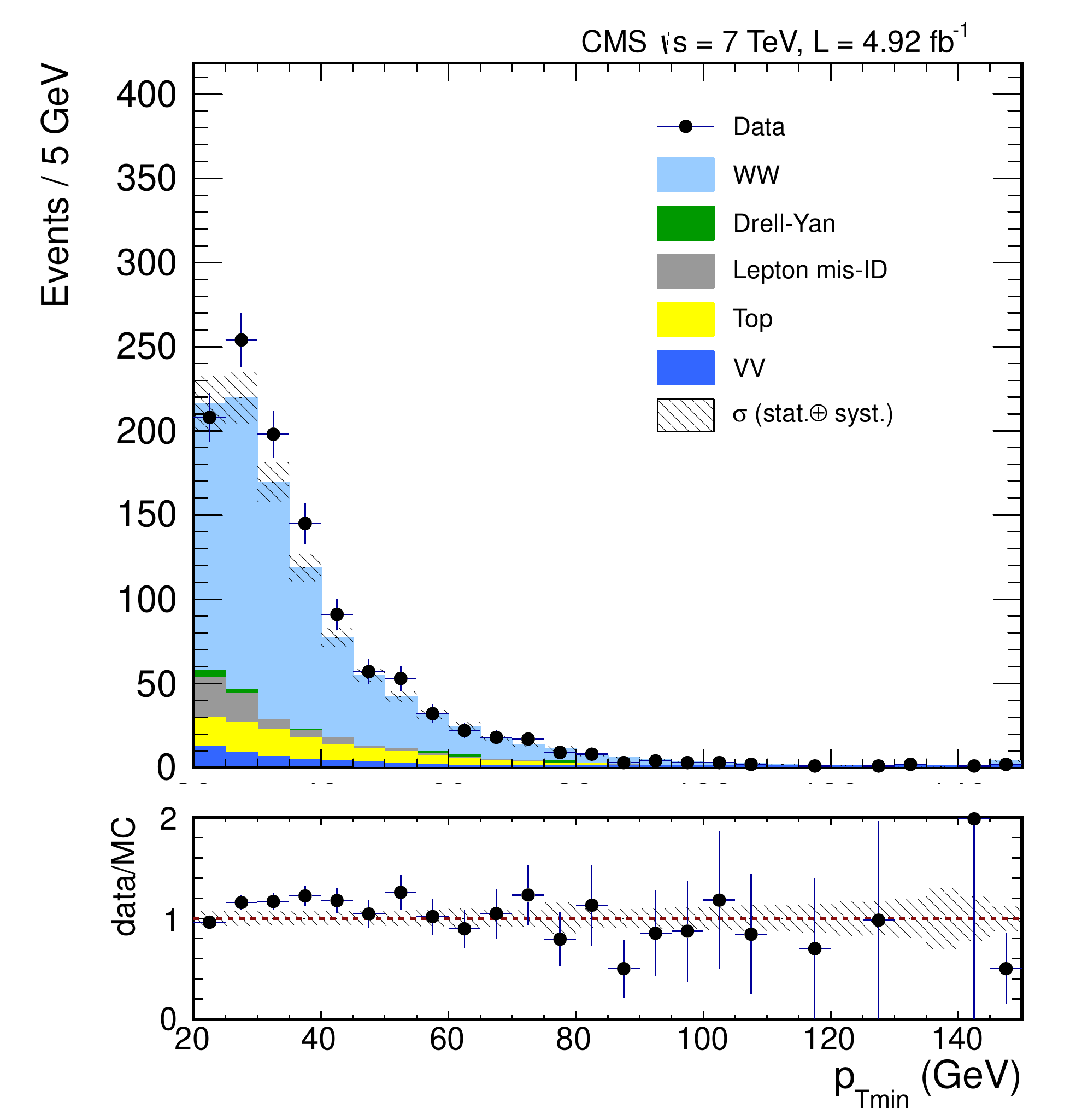}
 \includegraphics[width=0.48\textwidth]{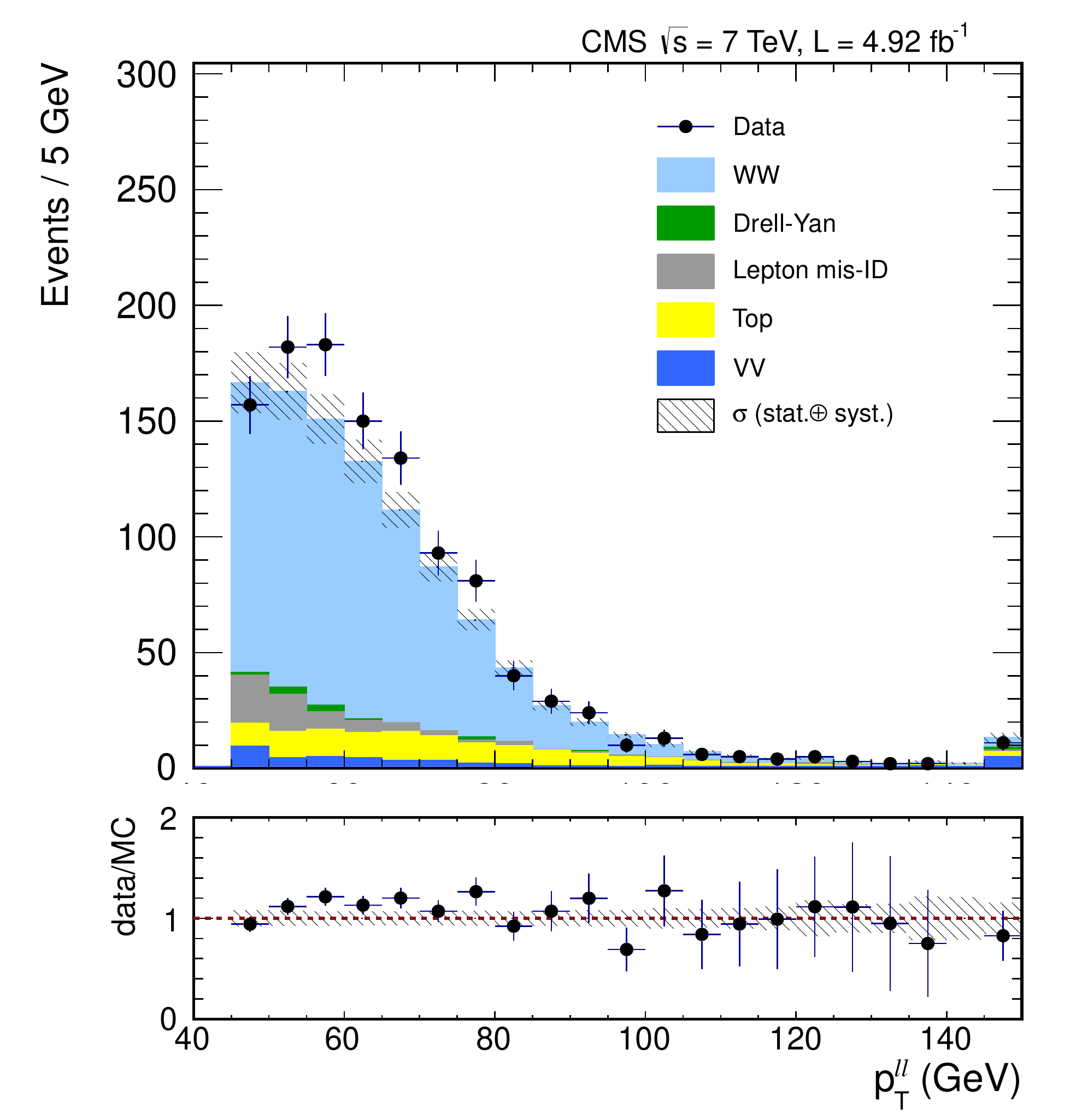}
 \includegraphics[width=0.48\textwidth]{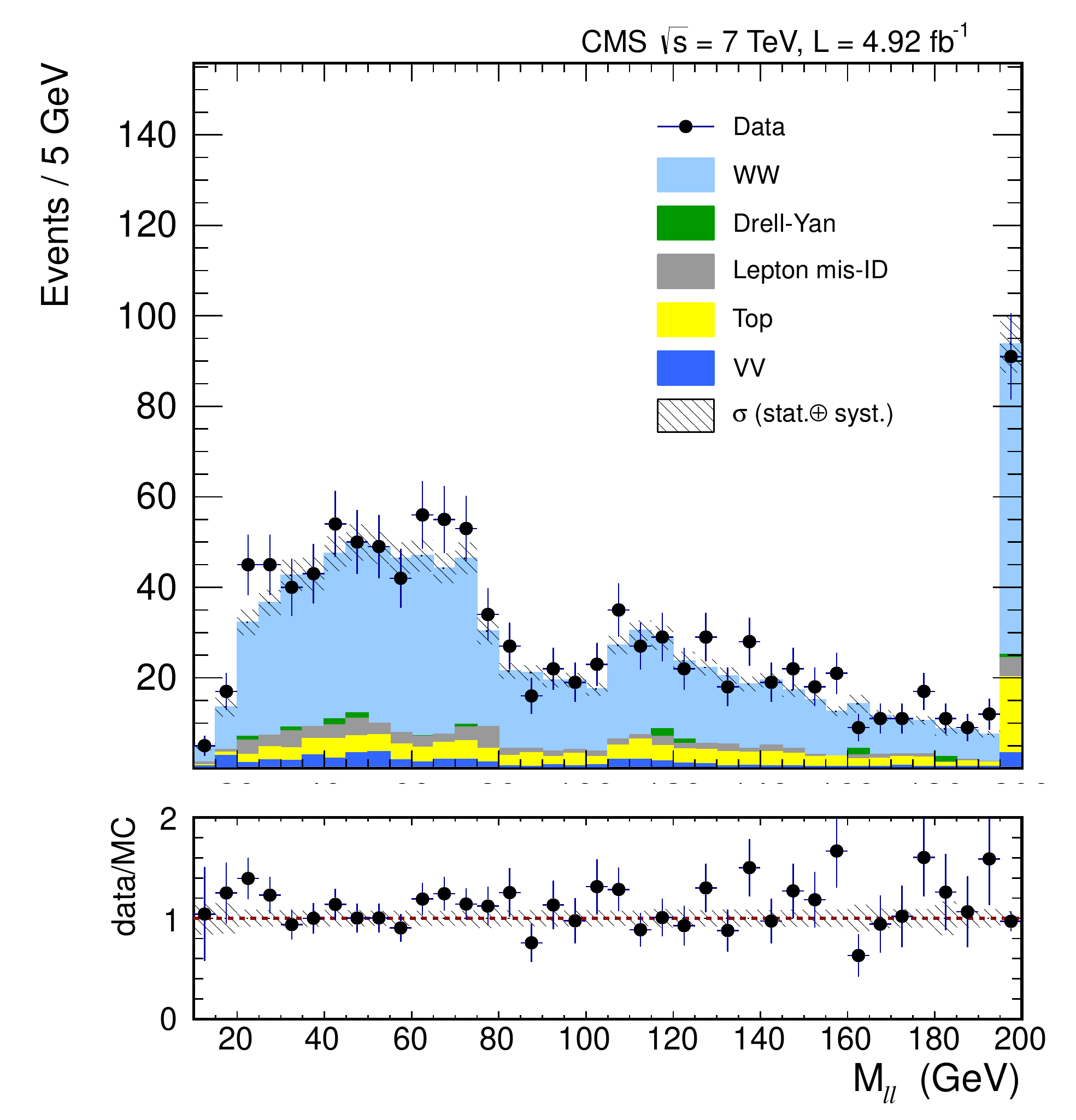}
 \caption{Distributions of the maximum lepton transverse momentum ($p_\mathrm{Tmax}$), the
minimum lepton transverse momentum ($p_\mathrm{Tmin}$), the dilepton transverse momentum
($\pt^{\ell\ell}$), and invariant mass (M$_{\ell\ell}$) at the final selection level.
Some of the backgrounds have been rescaled to the estimates based on control samples
in data, as described in the text. All leptonic channels are combined, and the uncertainty band
corresponds to the statistical and systematic uncertainties in the predicted yield.
The last bin includes the overflow. In the box below each distribution, the ratio of
the observed CMS event yield to the total SM prediction is shown.
}
\label{fig:distributions}
\end{center}
\end{figure*}

The estimated event yields for all processes after the event selection are summarised in
Table~\ref{tab:wwselection_all}. The distributions of the key analysis variables are
shown in Fig.~\ref{fig:distributions}.
\begin{table}[!ht]
  \begin{center}
    \topcaption{Signal and background predictions, compared to the yield in data.
    The prediction for the $\PWp\PWm$ process assumes the SM cross section value.
    \label{tab:wwselection_all}}
      \begin{tabular}{|c|c|}
        \hline
        Sample                           & Yield  $\pm$ stat. $\pm$ syst.  \\ \hline
        $\Pg\Pg \to \PWp\PWm$            & $46   \pm1   \pm14$             \\
        $\cPq\cPaq\to\PWp\PWm$       & $751  \pm4   \pm53$             \\ \hline
        \ttbar+tW                        & $129  \pm13  \pm20$             \\
        $\PW+\text{jets}$                         & $60   \pm4   \pm21$             \\
        WZ+ZZ                            & $29.4 \pm0.4 \pm2.0$            \\
        Z$/\Pgg^*\to\Pep\Pem$/$\Pgmp\Pgmm$ & $11.0 \pm5.1 \pm2.6$            \\
        $\PW\Pgg^{(*)}$                  & $18.8 \pm2.8 \pm4.7$            \\
        Z$/\Pgg^*\to\tau\tau$          & $0.0^{+1.0}_{-0.0}~^{+0.1}_{-0.0}$ \\ \hline

        Total Background                 & $247  \pm15  \pm30$             \\ \hline
        Signal + Background              & $1044 \pm15  \pm62$             \\ \hline \hline
        Data                             & $1134$                          \\ \hline
      \end{tabular}
  \end{center}
\end{table}

\section{Efficiencies and systematic uncertainties}
\label{sec:systematics}
The signal efficiency, which includes the acceptance of the detector, is estimated
using simulation and including both the $\cPq\cPaq\to\PWp\PWm$ and $\Pg\Pg\to
\PWp\PWm$ processes. Residual discrepancies in the lepton reconstruction and identification
efficiencies between data and simulation are corrected by determining data-to-simulation
scale factors measured using $\dyll$ events in the $\cPZ$ peak region~\cite{wzxs} that are
recorded with unbiased triggers. These factors depend on the lepton $\pt$ and $\abs{\eta}$
and are within 4\% (2\%) of unity for electrons (muons). Effects due to $\PW\to\tau\nu_{\tau}$
decays with $\tau$ leptons decaying into lower-energy electrons or muons are included in the
signal efficiency.

The experimental uncertainties in lepton reconstruction and identification efficiency,
momentum scale and resolution, $\MET$ modeling, and jet energy scale are applied to the
reconstructed objects in simulated events by smearing and scaling the relevant observables
and propagating the effects to the kinematic variables used in the analysis. A relative
uncertainty of 2.3\% in the signal efficiency due to multiple collisions within a bunch
crossing is taken from the observed variation in the efficiency in a comparison of two
different pileup scenarios in simulation, reweighted to the observed data.

The relative uncertainty in the signal efficiency due to variations in the PDFs and the
value of $\alpha_{s}$ is 2.3\% (0.8\%) for $\cPq\cPaq$ ($\Pg\Pg$) production, following the
{PDF4LHC} prescription~\cite{pdf4lhcInterim,pdf4lhcReport,ct10,MSTW2008pdf,NNPDFpdf,LHCHiggsCrossSectionWorkingGroup:2011ti}.
The effect of higher-order corrections, studied using the \textsc{mcfm} program~\cite{MCFM},
is found to be 1.5\% (30\%) for $\cPq\cPaq$ annihilation ($\Pg\Pg$) by varying the renormalisation
($\mu_R$) and factorisation ($\mu_F$) scales in the range ($\mu_0/2,2\mu_0$), with $\mu_0$ equal
to the mass of the \PW\ boson, and setting $\mu_R = \mu_F$. The $\PWp\PWm$ jet veto efficiency in data is
estimated from simulation and multiplied by a data-to-simulation scale factor derived from $\dyll$
events in the $\cPZ$ peak,
\begin{eqnarray*}
\epsilon_{\PWp\PWm}^\text{data} = \epsilon_{\PWp\PWm}^\mathrm{MC} \times \epsilon_{\cPZ}^\text{data}/\epsilon_{\cPZ}^\mathrm{MC},
\end{eqnarray*}
where $\epsilon_{\PWp\PWm}^\text{data}$ and $\epsilon_{\PWp\PWm}^\mathrm{MC}$ ($\epsilon_{\cPZ}^\text{data}$
and $\epsilon_{\cPZ}^\mathrm{MC}$) are the efficiencies for the jet veto on the $\PWp\PWm$ ($\cPZ$) process
for data and MC, respectively. The uncertainty in this efficiency is factorised into the uncertainty
in the $\cPZ$ efficiency in data and the uncertainty in the ratio of the $\PWp\PWm$ efficiency to the $\cPZ$
efficiency in simulation ($\epsilon_{\PWp\PWm}^\mathrm{MC}/\epsilon_{\cPZ}^\mathrm{MC}$). The former, which is
dominated by statistics, is 0.3\%. Theoretical uncertainties due to higher-order corrections contribute
most to the $\epsilon_{\PWp\PWm}^\mathrm{MC}/\epsilon_{\cPZ}^\mathrm{MC}$ ratio uncertainty, which is 4.6\%. The
data-to-simulation correction factor is close to unity, using the $\dyll$ events.

The uncertainties in the $\PW+\text{jets}$ and top-quark background predictions are evaluated to be 36\% and 18\%,
respectively, as described in Section~\ref{sec:backgrounds}. The total uncertainty in the $\dyll$ normalisation
is about 50\%, including both statistical and systematic contributions.

The theoretical uncertainties in the diboson cross sections are calculated by varying the renormalisation
and factorisation scales using the \textsc{mcfm} program~\cite{MCFM}. The effect of variations in the PDFs
and the value of $\alpha_{s}$ on the predicted cross section are derived by following the same prescription
as for the signal acceptance. Including the experimental uncertainties gives a systematic uncertainty of
around 10\% for WZ and ZZ processes. In the case of $\PW\Pgg^{(*)}$ backgrounds, it rises to 30\%, due
to the lack of knowledge of the overall normalisation. The total uncertainty in the background estimates is about
15\%, which is dominated by the systematic uncertainties in the normalisation of the top-quark and $\PW+\text{jets}$
backgrounds. A $2.2\%$ uncertainty is assigned to the integrated luminosity measurement~\cite{lumiPAS}.
A summary of the uncertainties is given in Table~\ref{tab:wwSystematics}.
For simplicity, averages of the estimates for WZ and ZZ backgrounds are shown.
\begin{table*}[!htb]
  \begin{center}
    \topcaption{Relative systematic uncertainties in the estimated signal and background
    yields, in units of percent.\label{tab:wwSystematics}}
      {\scriptsize
      \begin{tabular}{|l|c|c|c|c|c|c|c|c|c|}
        \hline
\multirow{10}{*}{} & $\cPq\cPaq$ & $\Pg\Pg$                & $\ttbar+\cPqt\PW$ & $\PW+\text{jets}$ & $\PW\cPZ$     & $\cPZ/\Pgg*$            & $\PW+\gamma$ & $\PW+\gamma^{*}$ & $\cPZ/\Pgg^*$ \\
                   & $\to \PWp\PWm$    & $\to \PWp\PWm$ &     &          & $+\cPZ\cPZ$    & $\to\ell\ell$ &             &                  & $\to \tau\tau$ \\

        \hline
        Luminosity                    & 2.2 & 2.2  & --    & --    & 2.2 & --    & 2.2  &  --   &  --   \\
        Trigger efficiency            & 1.5 & 1.5  & --    & --    & 1.5 & --    & 1.5  &  --   &  --   \\
        Lepton ID efficiency          & 2.0 & 2.0  & --    & --    & 2.0 & --    & 2.0  &  --   &  --   \\
        Muon momentum scale           & 1.5 & 1.5  & --    & --    & 1.5 & --    & 1.5  &  --   &  --   \\
        Electron energy scale         & 2.5 & 2.5  & --    & --    & 1.9 & --    & 2.0  &  --   &  --   \\
        $\MET$ resolution             & 2.0 & 2.0  & --    & --    & 2.0 & --    & 2.0  &  --   &  --   \\
        Jet veto efficiency           & 4.7 & 4.7  & --    & --    & 4.7 & --    & 4.7  &  --   &  --   \\
        Pileup                        & 2.3 & 2.3  & --    & --    & 2.3 & --    & 2.3  &  --   &  --   \\
        $\ttbar$+$\cPqt\PW$ normalisation     & --   & --    & 18   & --    & --   & --    & --    &  --   &  --   \\
       $\PW+\text{jets}$ normalisation         & --   & --    & --    & 36   & --   & --    & --    &  --   &  --   \\
       $\dyll$ normalisation          & --   & --    & --    & --    & --   & 50   & --    &  --   &  --   \\
       $\PW+\Pgg$ normalisation      & --   & --    & --    & --    & --   & --    & 30   &  --   &  --   \\
       $\PW+\Pgg^{*}$ normalisation  & --   & --    & --    & --    & --   & --    & --    & 30   &  --   \\
       $\dytt$ normalisation          & --   & --    & --    & --    & --   & --    & --    &  --   & 10   \\
        PDFs                          & 2.3 & 0.8  & --    & --    & 5.9 & --    & --    & --    &  --   \\
        Higher-order corrections      & 1.5 & 30   & --    & --    & 3.3 & --    & --    & --    &  --   \\
        \hline
      \end{tabular} }
  \end{center}
\end{table*}

\section{The WW cross section measurement}
\label{sec:results}
The number of events observed in the signal region is $N_{\text{data}} = 1134$.
The $\PWp\PWm$ yield is calculated by subtracting the expected contributions of the
various SM background processes, $N_\text{bkg} = 247\pm 15\stat
\pm 30\syst$ events. The inclusive cross section is obtained from the
expression
\begin{equation}
\sigma_{\PWp\PWm} = \frac{N_\text{data}-N_\text{bkg} }{\mathcal{L}_\text{int} \cdot \epsilon \cdot \left( 3 \cdot \mathcal{B}(\PW\to \ell\cPagn) \right)^2},
\label{eq:wwxs}
\end{equation}
where the signal selection efficiency $\epsilon$, including the detector acceptance
and averaging over all lepton flavours, is found to be $(3.28 \pm 0.02\stat \pm 0.26\syst)\%$
using simulation and taking into account the two production modes. As shown in Eq.~(\ref{eq:wwxs}),
the efficiency is corrected by the branching fraction for a W boson decaying to each lepton family,
$\mathcal{B}(\PW\to \ell\cPagn) = (10.80 \pm 0.09)\%$~\cite{pdg2012}, to estimate the final
inclusive efficiency for the signal.

The $\PWp\PWm$ production cross section in $\Pp\Pp$ collision data at $\sqrt{s} = 7\TeV$ is measured to be
\begin{equation*}
\sigma_{\PWp\PWm} = 52.4 \pm 2.0\stat\pm 4.5\syst\pm 1.2\lum\unit{pb}.
\end{equation*}
The statistical uncertainty is due to the total number of observed events. The systematic
uncertainty includes both the statistical component from the limited number of events and
systematic uncertainties in the background prediction, as well as the uncertainty in the
signal efficiency.

This measurement is consistent with the SM expectation of $47.0\pm2.0\unit{pb}$, based on
$\cPq\cPaq$ annihilation and gluon-gluon fusion. For the event selection used
in the analysis, the expected theoretical cross section may be larger by as much as 5\%
because of additional $\PWp\PWm$ production processes, such as diffractive production~\cite{Bruni:1993is},
double parton scattering, QED exclusive production~\cite{Pukhov:2004ca}, and Higgs boson
production with decay to $\PWp\PWm$. The dominant contribution of about 4\% would come from SM
Higgs production, assuming its mass to be near 125\GeV~\cite{HiggsPaper2012}.

The measured $\PWp\PWm$ cross section can be presented in terms of a ratio to the Z boson
production cross section in the same data set. The $\PWp\PWm$ to Z cross section ratio,
$\sigma_{\PWp\PWm}/\sigma_{\cPZ}$,  provides a good cross-check of this $\PWp\PWm$ cross
section measurement, using the precisely known Z boson production cross section as a
reference. This ratio has the advantage that some systematic effects cancel. More
precise comparisons between measurements from different data-taking periods are
possible because the ratio is independent of the integrated luminosity. The PDF
uncertainty in the theoretical cross section prediction is also largely cancelled
in this ratio, since both $\PWp\PWm$ and the Z boson are produced mainly via $\cPq\cPaq$
annihilation. The estimated theoretical value for this ratio is $[ 1.63 \pm 0.07\,\text{(theor.)} ]\times
10^{-3}$~\cite{wzxs}, where the scale uncertainty between both processes is considered
uncorrelated, while the PDF uncertainty is assumed fully correlated.

The $\cPZ$ boson production process is measured in the $\Pep\Pem$/$\Pgmp\Pgmm$ final states using
events passing the same lepton selection as in the $\PWp\PWm$ measurement and lying within the
$\cPZ$ mass window, where the purity of the sample is about 99.8\%~\cite{wzxs}. Nonresonant %
backgrounds (including $\dytt$) are estimated from $e\mu$ data, while the resonant component
of $\PW\cPZ$ and $\cPZ\cPZ$ processes is normalised to NLO cross sections using MC samples. Correlation of
theoretical and experimental uncertainties between the two processes is taken into account.
An additional 2\% uncertainty in the shape of the $\cPZ$ resonance due to final-state radiation
and higher-order effects is assigned. The latter is based on the difference between the
next-to-next-to-leading-order prediction from \textsc{fewz} 2.0~\cite{FEWZ} simulation code
and the MC generator used in the analysis, and on the renormalisation and factorisation scale
variation given by \textsc{FEWZ}.

The ratio of the inclusive $\PWp\PWm$ cross section to the Z cross section in the dilepton mass range
between 60 and 120\GeV is measured to be
\begin{equation*}
 \sigma_{\PWp\PWm}/\sigma_{\cPZ} = [ 1.79 \pm 0.16 (\text{stat.}{\oplus}\text{syst.}) ] \times 10^{-3},
\end{equation*}
in agreement with the theoretical expectation. The $\cPZ$ cross section resulting from this ratio,
assuming the standard model value for the $\PWp\PWm$ cross section, is 1.1\% higher than the inclusive
$\cPZ$ cross section measurement in CMS using the 2010 data set~\cite{wzxs}, which had an integrated
luminosity of 36\pbinv, but well within the systematic uncertainties of both measurements.

\section{Limits on the anomalous triple gauge-boson couplings}
\label{sec:aTGCs}
A search for anomalous TGCs is done using the effective Lagrangian approach
with the LEP parametrisation~\cite{LEPparametrization} without form factors.
The most general form of such a Lagrangian has 14 complex couplings (7 for
$\PW\PW\cPZ$ and 7 for $\PW\PW\Pgg$). Assuming electromagnetic gauge invariance and
charge and parity symmetry conservation, that number is reduced to five
real couplings:
$\Delta\kappa_Z$, $\Delta g_1^Z$, $\Delta\kappa_\gamma$, $\lambda_\cPZ$, and
$\lambda_\gamma$. Applying gauge invariance constraints leads to
\begin{align*}
\Delta\kappa_\cPZ =& \Delta g_1^\cPZ - \Delta\kappa_\gamma {\tan}^2(\theta_{\PW}), \\
\lambda_\cPZ      =& \lambda_\gamma,
\end{align*}
which reduces the number of independent couplings to three. In the SM, all
five couplings are zero. The coupling constants $\Delta g_1^\cPZ$ and $\Delta\kappa_\gamma$
parametrise the differences from the standard model values of 1 for both $g_1^\cPZ$ and $\kappa_\gamma$,
which are measures of the $\PW\PW\cPZ$ and $\PW\PW\Pgg$ coupling strengths, respectively.

The presence of anomalous TGCs would enhance the production rate for diboson
processes at high boson $\pt$ and high invariant mass. The effect of these
couplings is ascertained by evaluating the expected distribution of $p_\mathrm{Tmax}$,
the transverse momentum of the leading (highest-$\pt$) lepton, and by comparing it
to the measured distribution, using a maximum-likelihood fit. The $p_\mathrm{Tmax}$ is
a very sensitive observable for these searches, and it is widely used in the fully
leptonic final states, since the total mass of the event cannot be fully
reconstructed. The likelihood $L$ is defined as a product of Poisson probability
distribution functions for the observed number of events ($N_\text{obs}$) and
the combined one for each event, $P$($\pt$):
\begin{equation}
L = \re^{-N_\text{exp}}(N_\text{exp})^{N_\text{obs}} \prod_{i=1}^{N_\text{obs}} P(\pt),
\label{eq:likelihood}
\end{equation}
where $N_\text{exp}$ is the expected number of signal and background events. The
leading lepton $\pt$ distributions with anomalous couplings are simulated using
the \textsc{mcfm} NLO generator, taking into account the detector effects. The
distributions are corrected for the acceptance and lepton reconstruction efficiency,
as described in Section~\ref{sec:systematics}. The uncertainties in the quoted integrated
luminosity, signal selection, and background fraction are assumed to be Gaussian. These
uncertainties are incorporated in the likelihood function in Eq.~(\ref{eq:likelihood}) by
introducing nuisance parameters with Gaussian constraints. A set of points with nonzero
anomalous couplings is used and distributions between the points are extrapolated assuming a
quadratic dependence of the differential cross section as a function of the anomalous couplings.

\begin{figure}[!hbt]
  \begin{center}
    \includegraphics[width=0.48\textwidth]{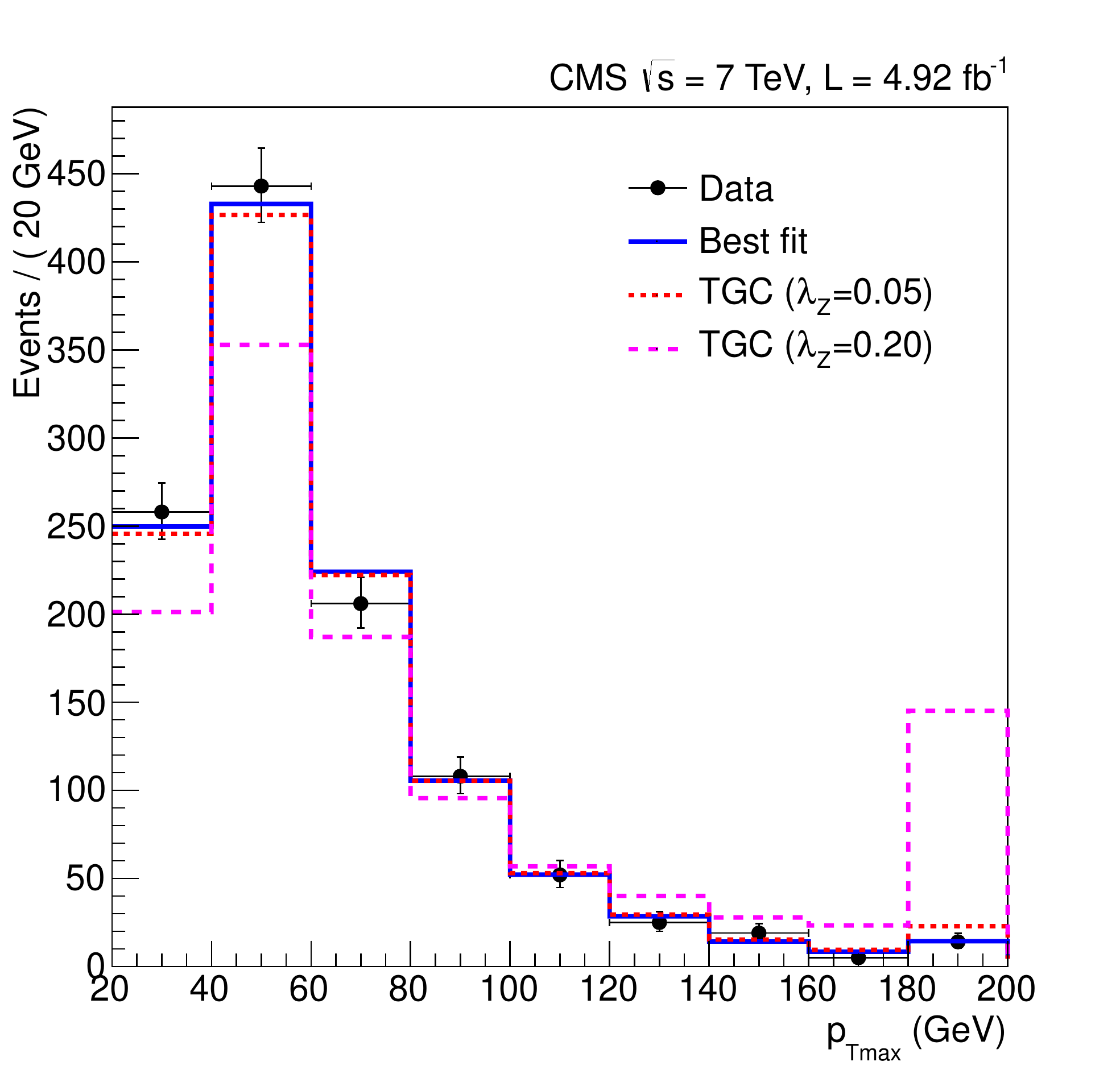}
    \caption[Leading lepton \pt{}]{Leading lepton \pt distribution in data (points with error bars)
    overlaid with the best fit using a two-dimensional $\lambda_{\cPZ}$-$\Delta g_1^{\cPZ}$ model (solid
    histogram) and two expected distributions with anomalous coupling value, $\lambda_{\cPZ} \neq 0$
    (dashed and dotted histograms). In the SM, $\lambda_{\cPZ} = 0$. The last bin includes the overflow.}
    \label{fig:atgc_pt}
  \end{center}
\end{figure}

Figure~\ref{fig:atgc_pt} shows the measured leading lepton \pt distributions in data and the predictions
for the SM $\PWp\PWm$ signal and background processes, as well as the expected distributions with
non-negative anomalous couplings, in the two-dimensional model $\lambda_{\cPZ}$-$\Delta g_1^{\cPZ}$.

\begin{figure}[hbt!]
    \includegraphics[width=0.49\textwidth]{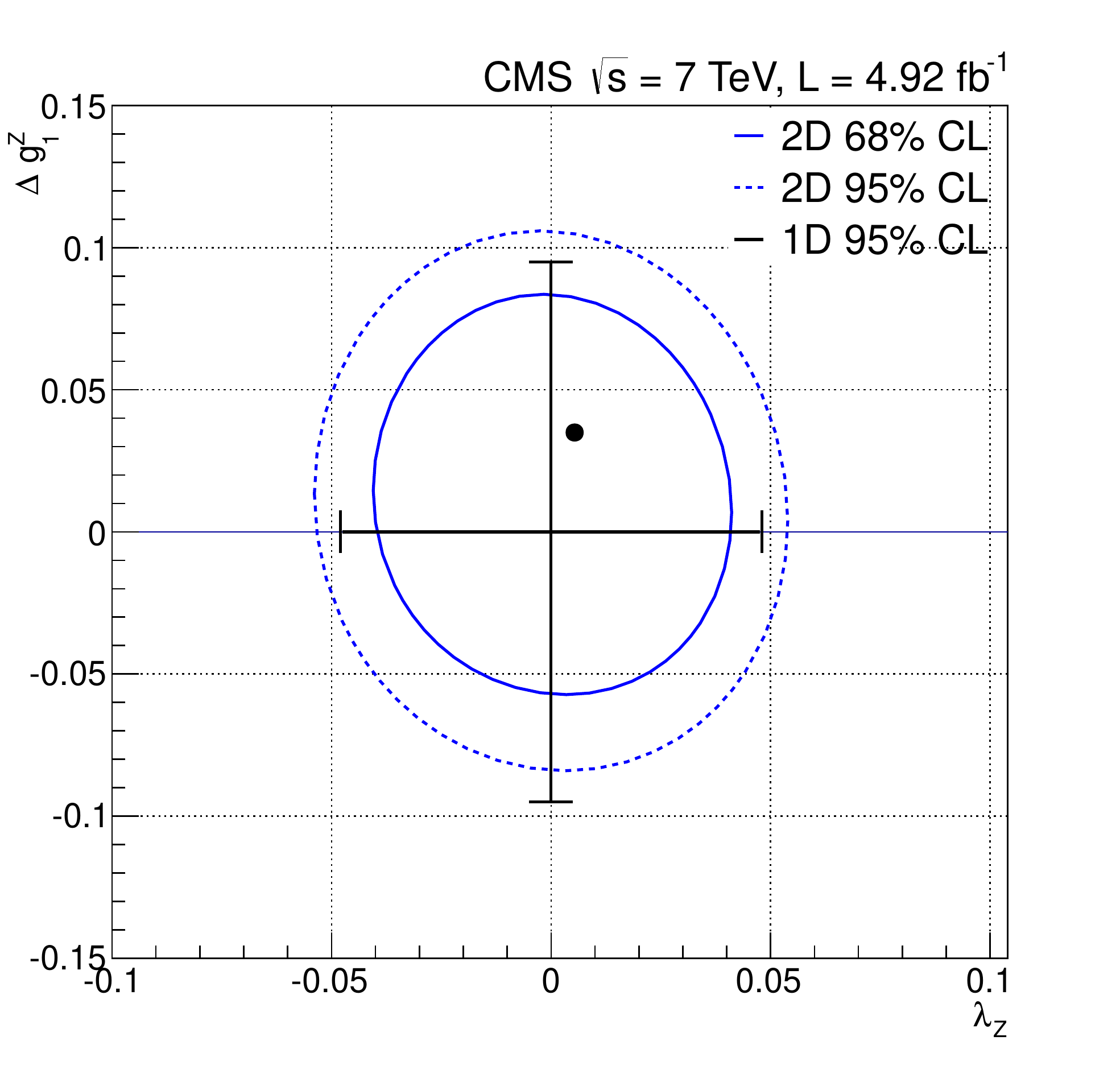}
    \includegraphics[width=0.49\textwidth]{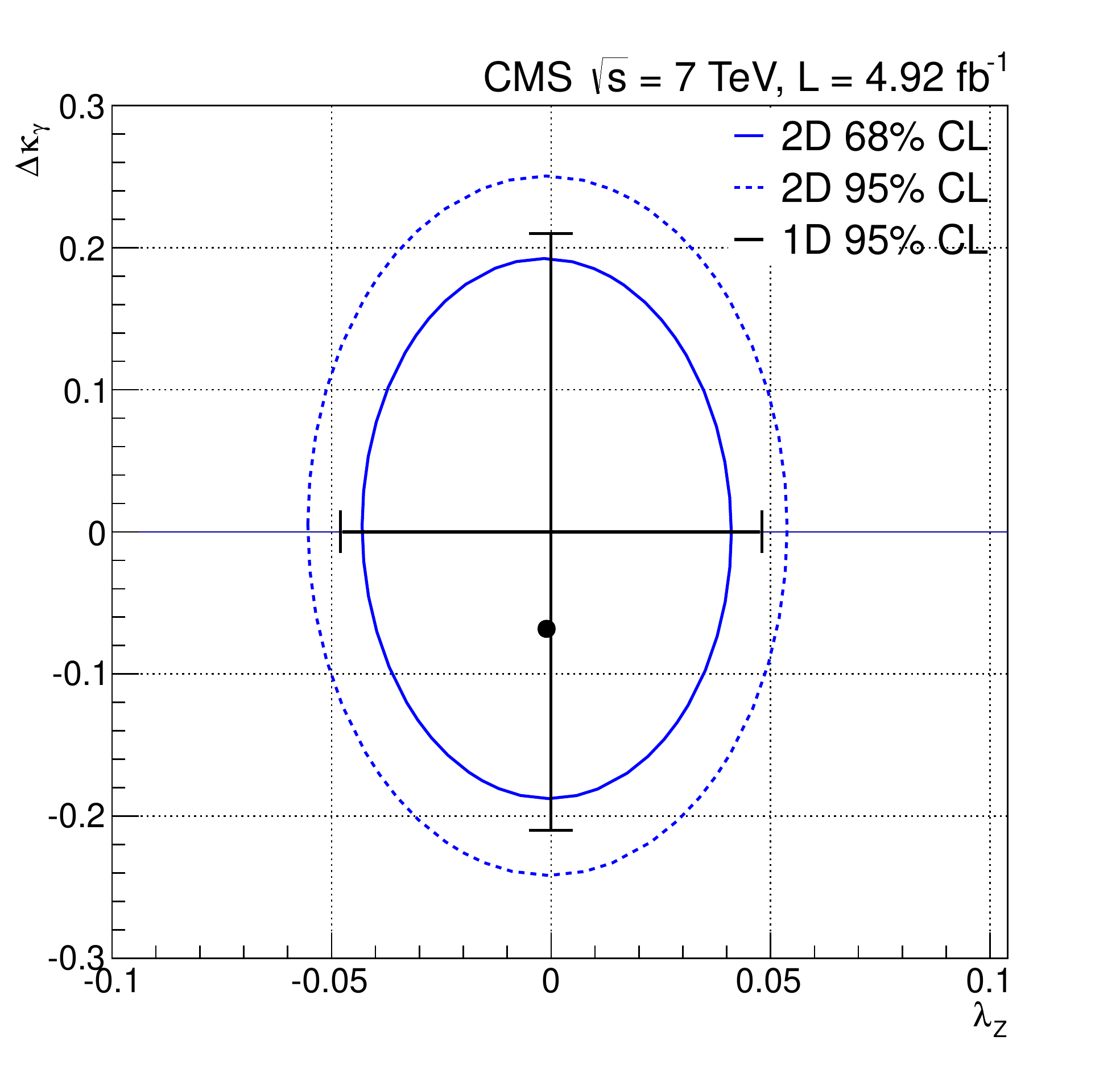}
  \caption{The 68\% (solid line) and 95\% CL (dashed line) limit contours, as well
    as the central value (point) of the fit results using unbinned fits, for $\Delta
    \kappa_\gamma = 0$ (\cmsLeft) and $\Delta g_1^{\cPZ} = 0$ (\cmsRight). The one-dimensional 95\% CL limit for
    each coupling is also shown.
  }
  \label{fig:contours}
\end{figure}

No evidence for anomalous couplings is found. The 95\% confidence level (CL) intervals of allowed
anomalous couplings values, setting the other two couplings to their SM expected values, are
\begin{align*}
-0.048 &\leq \lambda_{\cPZ} \leq 0.048,\\
-0.095 &\leq \Delta g^{\cPZ}_1 \leq 0.095,\\
-0.21 &\leq \Delta\kappa_\gamma \leq 0.22.
\end{align*}
The results presented here are comparable with the measurements performed by the ATLAS
Collaboration~\cite{atlaswwpaper2011} using the LEP parametrisation. These results are also comparable upon those
obtained at the Tevatron~\cite{TevatronTGCs,Tevatron2TGCs}, which are based on the HISZ parametrisation~\cite{HISZ} and LEP parametrisation with form factors, but they are not as precise as the combination of the LEP experiments~\cite{aleph:atgc,l3:atgc,opal:atgc}. Recently, CMS has set limits on these couplings ~\cite{CMSTGCs}, using different final state channel.
Our measurements clearly demonstrate that both the $\PW\PW\cPZ$ and $\PW\PW\Pgg$ couplings exist, as
predicted in the standard model ($g_1^{\cPZ} = 1$, $\kappa_\gamma = 1$). Figure~\ref{fig:contours}
displays the contour plots at the 68\% and 95\%~CL for the $\Delta\kappa_\gamma = 0$ and
$\Delta g_1^{\cPZ} = 0$ scenarios.

\section{Summary}
\label{sec:conclusions}
This paper reports a measurement of the $\PWp\PWm$ cross section in the
$\PWp\PWm\to\ell^{+}\nu\ell^{-}\cPagn$ decay channel in proton-proton
collisions at a centre of mass energy of 7\TeV, using the full CMS data set
of 2011. The $\PWp\PWm$ cross section is measured to be $52.4 \pm 2.0\stat\pm 4.5\syst\pm
1.2\lum\unit{pb}$, consistent with the NLO theoretical prediction,
$\sigma^\mathrm{NLO} (\Pp\Pp \to \PWp\PWm) = 47.0 \pm 2.0\unit{pb}$. No evidence
for anomalous $\PW\PW\cPZ$ and $\PW\PW\Pgg$ triple gauge-boson couplings is found,
and stringent limits on their magnitude are set.

\section*{Acknowledgements}
We congratulate our colleagues in the CERN accelerator departments for the excellent performance of the LHC and thank the technical and administrative staffs at CERN and at other CMS institutes for their contributions to the success of the CMS effort. In addition, we gratefully acknowledge the computing centres and personnel of the Worldwide LHC Computing Grid for delivering so effectively the computing infrastructure essential to our analyses. Finally, we acknowledge the enduring support for the construction and operation of the LHC and the CMS detector provided by the following funding agencies: BMWF and FWF (Austria); FNRS and FWO (Belgium); CNPq, CAPES, FAPERJ, and FAPESP (Brazil); MEYS (Bulgaria); CERN; CAS, MoST, and NSFC (China); COLCIENCIAS (Colombia); MSES (Croatia); RPF (Cyprus); MoER, SF0690030s09 and ERDF (Estonia); Academy of Finland, MEC, and HIP (Finland); CEA and CNRS/IN2P3 (France); BMBF, DFG, and HGF (Germany); GSRT (Greece); OTKA and NKTH (Hungary); DAE and DST (India); IPM (Iran); SFI (Ireland); INFN (Italy); NRF and WCU (Republic of Korea); LAS (Lithuania); CINVESTAV, CONACYT, SEP, and UASLP-FAI (Mexico); MSI (New Zealand); PAEC (Pakistan); MSHE and NSC (Poland); FCT (Portugal); JINR (Armenia, Belarus, Georgia, Ukraine, Uzbekistan); MON, RosAtom, RAS and RFBR (Russia); MSTD (Serbia); SEIDI and CPAN (Spain); Swiss Funding Agencies (Switzerland); NSC (Taipei); ThEPCenter, IPST and NSTDA (Thailand); TUBITAK and TAEK (Turkey); NASU (Ukraine); STFC (United Kingdom); DOE and NSF (USA).

Individuals have received support from the Marie-Curie programme and the European Research Council and EPLANET (European Union); the Leventis Foundation; the A. P. Sloan Foundation; the Alexander von Humboldt Foundation; the Belgian Federal Science Policy Office; the Fonds pour la Formation \`a la Recherche dans l'Industrie et dans l'Agriculture (FRIA-Belgium); the Agentschap voor Innovatie door Wetenschap en Technologie (IWT-Belgium); the Ministry of Education, Youth and Sports (MEYS) of Czech Republic; the Council of Science and Industrial Research, India; the Compagnia di San Paolo (Torino); the HOMING PLUS programme of Foundation for Polish Science, cofinanced by EU, Regional Development Fund; and the Thalis and Aristeia programmes cofinanced by EU-ESF and the Greek NSRF.
\bibliography{auto_generated}   
\cleardoublepage \appendix\section{The CMS Collaboration \label{app:collab}}\begin{sloppypar}\hyphenpenalty=5000\widowpenalty=500\clubpenalty=5000\textbf{Yerevan Physics Institute,  Yerevan,  Armenia}\\*[0pt]
S.~Chatrchyan, V.~Khachatryan, A.M.~Sirunyan, A.~Tumasyan
\vskip\cmsinstskip
\textbf{Institut f\"{u}r Hochenergiephysik der OeAW,  Wien,  Austria}\\*[0pt]
W.~Adam, T.~Bergauer, M.~Dragicevic, J.~Er\"{o}, C.~Fabjan\cmsAuthorMark{1}, M.~Friedl, R.~Fr\"{u}hwirth\cmsAuthorMark{1}, V.M.~Ghete, N.~H\"{o}rmann, J.~Hrubec, M.~Jeitler\cmsAuthorMark{1}, W.~Kiesenhofer, V.~Kn\"{u}nz, M.~Krammer\cmsAuthorMark{1}, I.~Kr\"{a}tschmer, D.~Liko, I.~Mikulec, D.~Rabady\cmsAuthorMark{2}, B.~Rahbaran, C.~Rohringer, H.~Rohringer, R.~Sch\"{o}fbeck, J.~Strauss, A.~Taurok, W.~Treberer-Treberspurg, W.~Waltenberger, C.-E.~Wulz\cmsAuthorMark{1}
% ...
\vskip\cmsinstskip
% ...
1:~~Also at Vienna University of Technology, Vienna, Austria\\
2:~~Also at CERN, European Organization for Nuclear Research, Geneva, Switzerland\\
% ...

\end{sloppypar}
\end{document}